\documentclass[aps, prx, twocolumn, showpacs, floatfix, superscriptaddress]{revtex4-2}
\usepackage{graphicx,amsmath,amssymb,physics,bm,bbm}
\graphicspath{{images/}}
\begin{document}

\title{Conditions for Equivalent Noise Sensitivity of Geometric and Dynamical Quantum Gates}
\author{R.~K.~L.~Colmenar}
\email{ralphkc1@umbc.edu}
\affiliation{Department of Physics, University of Maryland Baltimore County, Baltimore, Maryland 21250, USA}
\author{Utkan G\"{u}ng\"{o}rd\"{u}}
\affiliation{Laboratory for Physical Sciences, College Park, Maryland 20740, USA}
\affiliation{Department of Physics, University of Maryland, College Park, Maryland 20742, USA}
\affiliation{Department of Physics, University of Maryland Baltimore County, Baltimore, Maryland 21250, USA}
\author{J.~P.~Kestner}
\affiliation{Department of Physics, University of Maryland Baltimore County, Baltimore, Maryland 21250, USA}

\begin{abstract}
Geometric quantum gates are often expected to be more resilient than dynamical gates against certain types of error, which would make them ideal for robust quantum computing. However, this is still up for debate due to seemingly conflicting results in the literature. Here we use dynamical invariant theory in conjunction with filter functions in order to analytically characterize the noise sensitivity of an arbitrary quantum gate. For any control Hamiltonian that produces a geometric gate, we find that under certain common conditions one can construct another control Hamiltonian that produces an equivalent dynamical gate with identical noise sensitivity (as characterized by the filter function). Our result holds for a Hilbert space of arbitrary dimensions, but we illustrate our result by examining experimentally relevant single-qubit scenarios and providing explicit examples of equivalent geometric and dynamical gates.
\end{abstract}

\maketitle

\section{INTRODUCTION}
\label{sec:intro}
One of the biggest roadblocks in quantum computing is developing techniques that enable control of quantum information under a certain error threshold~\cite{Knill_2005}. Among the plethora of potential candidates for robust quantum control, geometric quantum computation (GQC)~\cite{Ekert_2000,Sjoqvist_2016,Zhang_2021} stands out owing to its elegant formulation in terms of concepts from differential geometry and topology. Put simply, a geometric quantum gate is a type of quantum gate for which it is possible to attribute a geometric interpretation to the accumulated phase. The usual paradigm is to generate a desired quantum gate in a basis of cyclic states. After an adiabatic~\cite{Berry_1984} or nonadiabatic~\cite{Aharonov_1987} cyclic evolution, these states accumulate a phase that depends on the qubit's spectrum. If the computational basis is encoded in an energetically nondegenerate (degenerate) subspace of the total Hilbert space, the computational basis accumulates an Abelian (non-Abelian) phase~\cite{Anandan_1988,Zanardi_1999}. This phase can be decomposed into a dynamical and a geometric component. A geometric gate is naturally produced when the dynamical component of the total phase is trivial, though that condition is not necessary~\cite{Zhu_2003}. Further extension to noncyclic evolution has also been made~\cite{Kult_2006}. Experimentally, geometric gates have been realized in nuclear magnetic resonance~\cite{Jones_2000, Du_2006, Ota_2009_EXP}, trapped-ion~\cite{Leibfried_2003,Guo_2020}, solid-state~\cite{Zu_2014, Wang_2016,Sekiguchi_2017,Ishida_2018,Nagata_2018,Huang_2019}, and superconducting qubits~\cite{Tan_2014, Song_2017,Xu_2018,Xu_2020,Abdumalikov_2013}. In certain cases, the underlying topology of the Hilbert space can give rise to a topological geometric phase~\cite{Kitaev_2003,Nayak_2008,Sarma_2015}. For instance, whereas a nontopological adiabatic geometric phase depends on the cyclic path taken in the system's parameter space, a topological one depends only on whether the path encloses a nontrivial topological feature (e.g., a hole or a cut in parameter space). Such topological phases can emerge in 5/2-fractional quantum Hall systems~\cite{Sarma_2005} or semiconductor nanowire structures hosting Majorana zero modes ~\cite{Lutchyn_2010,Oreg_2010}. In this work, we restrict our considerations only to conventional systems with nontopological type of geometric phase.

The primary motivation for using GQC is the expectation that geometric gates are intrinsically more robust than dynamical gates. More rigorously, geometric gates are known to be robust against noise that only affects the rate at which the cyclic evolution path is traversed but not necessarily against noise that deforms the cyclic path~\cite{Dong_2021}. However, geometric gates are often considered to be generally superior to dynamical gates by the reasoning that, since geometric phase is a global feature of quantum evolution, it must be intrinsically resilient to noise that only generates local perturbations in the system's evolution path~\cite{Pachos_2001,Zhang_2021}. Thus, a majority of the effort on GQC focuses on finding experimentally feasible ways of eliminating dynamical phase contributions in a gate. Numerous studies on geometric gates, both theoretical~\cite{Ekert_2000, Carollo_2003, DeChiara_2003, Zhu_2005, DeChiara_2007, Wang_2007, Thomas_2011, Liang_2016, Chen_2018, Liu_2019, Chen_2020,Pachos_2001,Dong_2021} and experimental~\cite{Berger_2013, Kleisler_2018, Xu_2020}, have shown evidence to support the robustness claim. However, there are also studies that report control situations in which geometric gates are not intrinsically more robust than dynamical gates~\cite{Nazir_2002,Blais_2003,Ota_2009,Zheng_2016,Dong_2021} and, in certain scenarios, their sensitivity to noise deteriorates~\cite{Solinas_2004, Carollo_2004, Zhu_2005, Dajka_2007, Johansson_2012}.

Here we investigate the robustness of geometric and dynamical gates against coherent control parameter noise. We use dynamical invariant theory~\cite{Lewis_1967} in conjunction with filter functions~\cite{Green_2013, Ball_2015} to analytically characterize how noise sensitivity changes with the type of accumulated phase. We show that for any geometric gate it is possible to find, under certain common conditions, an equivalent gate with the same filter function but with a phase whose nature can be continuously varied from purely geometric to purely dynamical. In other words, we show within our framework that noise robustness and phase type are independent properties. This contradicts any naive expectation that geometric gates should generally be superior to dynamical gates. Our analysis applies equally to adiabatic and non-adiabatic geometric gates. We explicitly demonstrate our result in experimentally relevant single-qubit cases, including both Abelian and non-Abelian geometric gates. We also discuss how the presence of control constraints can break this equivalence of geometric and dynamical robustness capability and give rise to preferential phase robustness. Our result may reconcile decades of seemingly contradictory claims on geometric gate robustness within the literature. Furthermore, our result calls into question the primary motivation for using GQC.

\begin{figure*}
    \centering
    \includegraphics[scale=0.32]{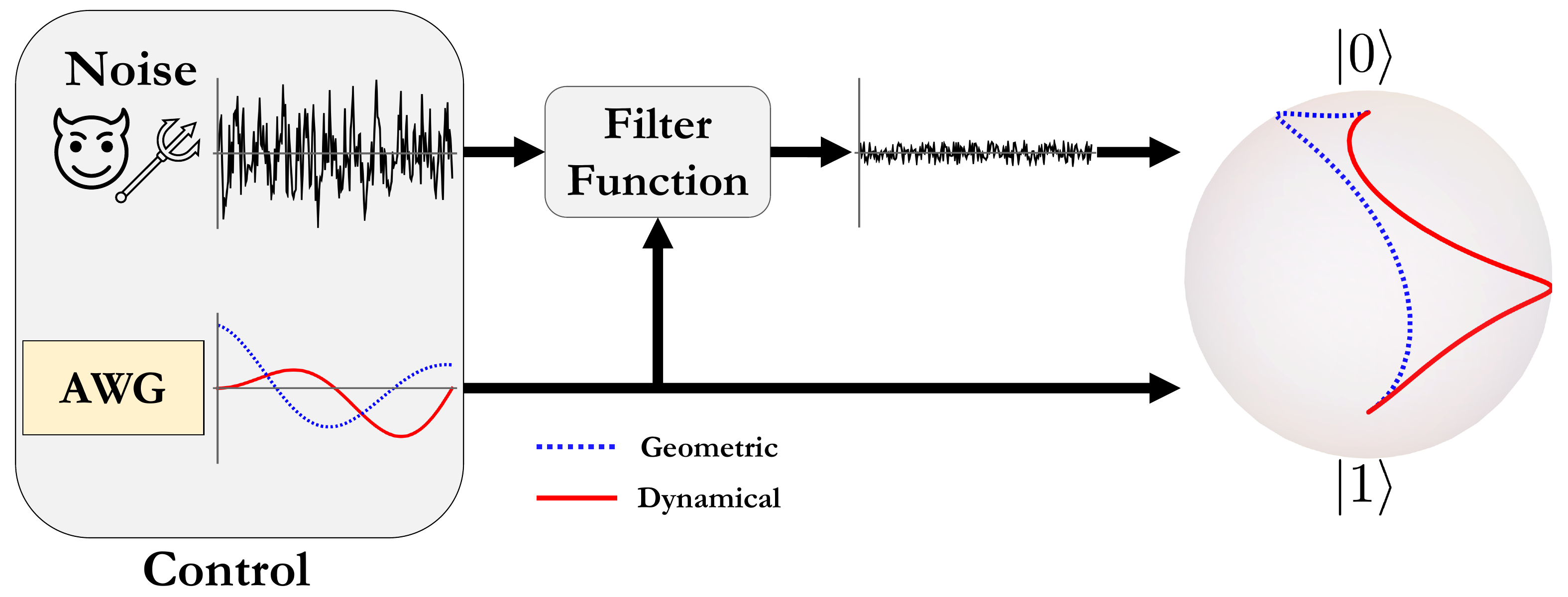}
    \caption{A schematic diagram summarizing our main result. The qubit's state is manipulated by applying control fields, e.g. using an arbitrary waveform generator (AWG), that are subject to some noise process (schematically represented here by a demon). The control fields determine an object called a filter function that characterizes the control's sensitivity to noise. In this diagram, the control is robust since it suppresses the effects of noise. Geometric gates are often expected to be more robust against noise than dynamical gates due to the geometric properties of their phase. Our main result shows that, for common types of noise in the absence of control constraints, one can actually construct different control fields that produce evolution paths with accrued phases ranging from purely geometric to purely dynamical that all result in the same gate and noise sensitivity.}
    \label{fig:overview}
\end{figure*}

\section{THEORY}
\label{sec:theory}
\subsection{Dynamical invariants}
\label{subsec:DI}
We begin by briefly describing how a geometric gate is generated. We temporarily restrict our attention to the Abelian case. A natural framework for considering geometric phase is through the theory of dynamical invariants~\cite{Lewis_1967,Mostafazadeh_2001}. Although dynamical invariants have previously been used in the context of quantum control~\cite{Chen_2011,Torrontegui_2011,Chen_2012,Gungordu_2012,Torrontegui_2014,Levy_2018,Guery-Odelin_2019}, in this work we only use them as a convenient way to describe the dynamical and geometric phases. Consider a qubit system whose evolution is governed by some Hamiltonian $H(t)$. A dynamical invariant $I(t)$ is a solution to the Liouville-von Neumann equation
\begin{equation}
\label{eq:liouville}
i \frac{\partial I(t)}{\partial t} - \left[H(t),I(t)\right] = 0,
\end{equation}
where we use units such that $\hbar=1$. The eigenvectors $\ket{\phi_n(t)}$ of $I(t)$ are related to the solutions of the Schr\"{o}dinger equation by a global phase factor, i.e., $\ket{\psi_n(t)} = \mathrm{e}^{i\alpha_n(t)}\ket{\phi_n(t)}$, where $\alpha_n(t)$ are the Lewis-Riesenfeld phases given by \cite{Lewis_1969}
\begin{gather}
\label{eq:lewis-riesenfeld-phase}
\alpha_n(t) = \alpha_{n,g}(t) + \alpha_{n,d}(t), \\
\label{eq:geometric-phase}
\alpha_{n,g}(t) = \int_{0}^{t} \ev**{i\partial_{t^\prime}}{\phi_{n}(t^\prime)} \mathrm{d}t^\prime, \\
\label{eq:dynamical-phase}
\alpha_{n,d}(t) = -\int_{0}^{t} \ev**{H(t^\prime)}{\phi_{n}(t^\prime)} \mathrm{d}t^\prime,
\end{gather}
with the subscripts $g$ and $d$ denoting the geometric and dynamical phases, respectively. We fix the $U(1)$ gauge freedom on our choice of $\ket{\phi_n(t)}$ by setting $\ket{\phi_n(0)}= \ket{\phi_n(T)}$, where $T$ is the gate time. This particular choice is consistent with Berry's adiabatic geometric phase \cite{Berry_1984} and generalizations thereof \cite{Wilczek_1984,Aharonov_1987,Anandan_1988}. One can show that, unlike the dynamical phase that generally depends on $T$, the geometric phase is independent of $T$ and is completely determined by the underlying geometric/topological property of the evolution path in Hilbert space. Within this framework, the evolution operator $U(t)$ can be expressed as
\begin{equation}
\label{eq:U_c}
U(t) = \sum_{n} \mathrm{e}^{i \alpha_n(t)} \ket{\phi_n(t)}\bra{\phi_n(0)}.
\end{equation}
A geometric gate is produced if the final accumulated dynamical phase is trivial, which can be ensured by, for example, carefully choosing the Hamiltonian so that the integral in Eq.~\eqref{eq:dynamical-phase} vanishes or by using composite pulses \cite{Ichikawa_2012}. 

\subsection{Filter functions}
\label{subsec:FF}
The robustness of a gate can be quantified using filter functions \cite{Green_2013, Ball_2015}, which provide a convenient method of quantifying the gate fidelity's susceptibility to noise of a given spectral composition. We limit our analysis to first-order filter functions, which means that we only consider weak and wide-sense stationary noise processes. A noisy $\mathfrak{su}(N)$ Hamiltonian can be decomposed as 
\begin{equation}\label{eq:Htot}
H(t) = H_{c}(t) + H_{e}(t),
\end{equation}
where $H_c(t)$ is the ideal deterministic control Hamiltonian and $H_e(t)$ is the stochastic error Hamiltonian, 
\begin{equation}\label{eq:H}
H_c(t) = \bm{h}_c(t)\cdot\bm{\sigma}, \quad H_e(t) = \sum_{q} \delta_q(t) \bm{\chi}_q\left[\bm{h}_c(t)\right]\cdot \bm{\sigma},
\end{equation}
where $q$ indexes a set of uncorrelated stochastic variables $\delta_q(t)$, $\bm{\chi}_q$ is the vector describing the first-order sensitivity of the control Hamiltonian to $\delta_q(t)$, and $\bm{\sigma}$ is a vector comprising the $N^2-1$ traceless Hermitian generators of $\mathfrak{su}(N)$.
Most commonly the sensitivity vector $\bm{\chi}_q$ is of the general linear form
\begin{equation}\label{eq:chi}
    \bm{\chi}_q\left[\bm{h}_c (t)\right] = \bm{a}_q + M_q(t)\bm{h}_c(t),
\end{equation}
where $\bm{a}_q$ is independent of the control (i.e., additive noise) and $M_q$ is likewise a real matrix accounting for sensitivity linearly proportional to some subset of the control (e.g., multiplicative noise). We assume this form for the remainder of the paper.

For sufficiently weak noise, we can compactly express the ensemble-averaged gate infidelity as
\begin{equation}
\label{eq:avg-infidelity}
\expval{\mathcal{I}} \approx \frac{1}{2\pi} \sum_{q} \int_{-\infty}^{\infty} \mathrm{d}\omega \, S_{q}(\omega)F_{q}(\omega),
\end{equation}
where $S_q(\omega)$ denotes the power spectral density for the stochastic variable $\delta_q(t)$ and $F_q(\omega)$ is the corresponding filter function. This is true only when the Magnus expansion of the evolution operator converges and higher-order noise contributions to the average gate infidelity are negligible \cite{Green_2013}. Fortunately, these conditions can be easily satisfied in a well-prepared system such as in many state-of-the-art quantum devices that routinely achieve gate fidelities above 99\%. Thus, it is safe to focus only on the first-order term.

Denote the $N\times N$ unitary evolution operator generated by the control Hamiltonian, $H_c$, in the absence of noise as the time-ordered exponential
\begin{equation}
    U_c(t) = \mathcal{T} e^{-i \int_0^t \mathrm{d}t^\prime H_c(t^\prime)} \equiv e^{-i\bm{\theta}(t)\cdot\bm{\sigma}/2}.
\end{equation}
We can also represent $U_c$ via its adjoint representation, $R$, defined through
\begin{equation}
\label{eq:adjoint-rep}
U_c\left(\bm{x}\cdot\bm{\sigma}\right)U_{c}^{\dagger} \equiv \left(R\bm{x}\right)\cdot \bm{\sigma} \implies R_{ij} = \tr(\sigma_i U_c \sigma_j U_c^{\dagger})/N.
\end{equation}
For example, in the case of $N=2$,
$R(t) = \exp\left(\bm{\theta}(t)\cdot \bm{L}\right)$,
where $\bm{L}$ is the vector of generators of $\mathfrak{so}(3)$ isomorphic to $\bm{\sigma}$. In general, the filter function can be interpreted geometrically as the magnitude of a complex vector 
\begin{gather}
\label{eq:filter-function}
F_q(\omega) = \bm{R}(\omega) \cdot \bm{R}(\omega)^*, \\
\label{eq:R-freq}
\bm{R}(\omega) = \int_{0}^{T} R^\intercal(t)\bm{\chi}_q\left[\bm{h}_c(t)\right] \mathrm{e}^{-i \omega t} \mathrm{d}t.
\end{gather}

\begin{figure*}
    \centering
    \includegraphics[scale=0.7]{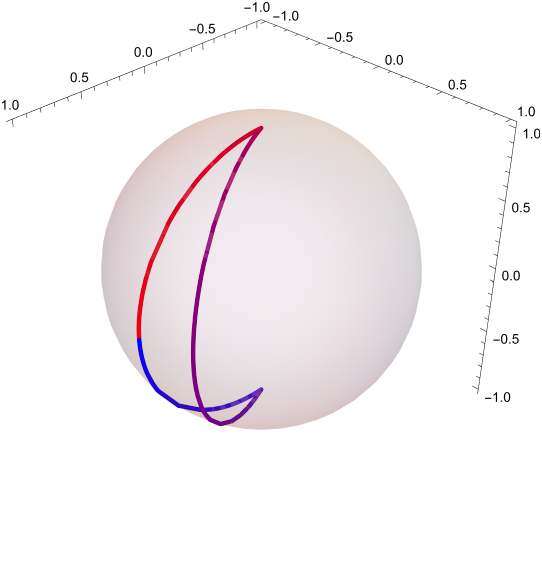}
    \includegraphics[scale=0.7]{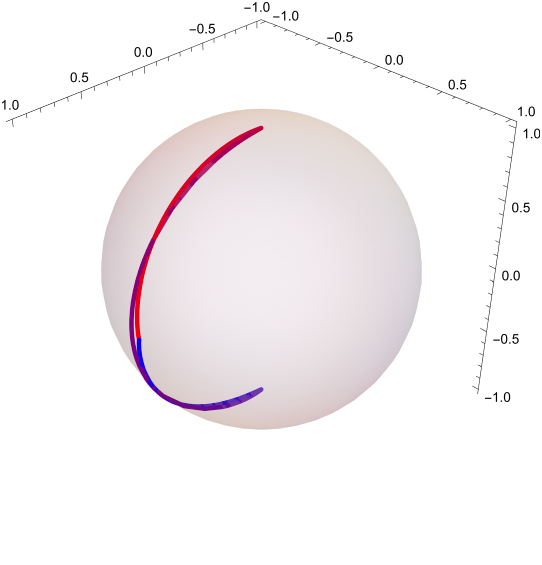}
    \caption{An illustration of a dynamical invariant eigenvector's evolution along the Bloch sphere for the geometric $X_{\frac{\pi}{2}}$ gate (left) and the dynamical $X_{\frac{\pi}{2}}$ gate (right). The path is traversed twice and its orientation is determined by the color gradient that begins with red and ends with blue. We see that $\ket{\phi_+(t)}$ traces out a loop with nonzero area in the geometric case. In contrast, the loop in the dynamical case encloses zero area.}
    \label{fig:path}
\end{figure*}

\subsection{Geometric and dynamical gates with identical filter functions}
\label{subsec:main-result}
For any control scheme $\bm{h}_c(t)$ that generates a gate geometrically, we now derive conditions under which there exists a different control $\bm{\tilde{h}_c}(t)$ that generates the same gate dynamically \emph{with an identical filter function}.  To this end, we calculate the filter function of two different control Hamiltonians, $H_c(t)$ and $\tilde{H}_c(t)$. Since for any arbitrary pair of time-dependent Hermitian operators $H_c$ and $\tilde{H}_c$ one can define a unitary $V = V_1 V_2$ where $\dot{V_1} = -i \tilde{H}_c V_1$ and $\dot{V_2} = +i H_c V_2$, the two can be related without loss of generality via a quantum canonical transformation 
\begin{equation}\label{eq:transformation}
\tilde{H}_c = V H_c V^\dagger - iV \dot{V}^\dagger.    
\end{equation}  
Note that we are not invoking a frame transformation here---indeed, the whole Hamiltonian is not subject to this transformation, only the control part---we are only using a convenient mathematical way to encapsulate in $V$ the difference between any two control Hamiltonians within the same frame. The geometric and dynamical phases produced by the two different control fields differ by a shift that is easily expressed in terms of $V$ \cite{Wei_1995,Mostafazadeh_2001}:
\begin{gather}
\label{eq:geom-phase-change}
\tilde{\alpha}_{n,g}(t) = \alpha_{n,g}(t) + \int_{0}^{t} \ev**{i V^\dagger \dot{V}}{\phi_{n}(t^\prime)} \mathrm{d}t^\prime, \\
\label{eq:dyn-phase-change}
\tilde{\alpha}_{n,d}(t) = \alpha_{n,d}(t) - \int_{0}^{t} \ev**{i V^\dagger \dot{V}}{\phi_{n}(t^\prime)} \mathrm{d}t^\prime.
\end{gather}
We emphasize again that, although it is well known \cite{Bitter_1987, Giavarini_1989, Wei_1995} that a frame transformation leaves the total Lewis-Riesenfeld phase invariant while shifting the dynamical and geometric phases, we are merely noting that different control Hamiltonians within a fixed frame generally have different dynamical and geometric phases and the difference is elegantly quantified in terms of the relationship $V$ between the Hamiltonians. As we return to below, the noise model of Eq.~\eqref{eq:chi} remains fixed, as it must in a fixed frame.

The control evolution operator induced by $\tilde{H}_c$ can be written in terms of that induced by $H_c$ as $\tilde{U}_c(t) = V(t) U_c(t) V^{\dagger}(0)$. Likewise, the adjoint representations of the evolutions can be related, denoting the adjoint representation of $V$ as $Q$, as
\begin{equation}
\label{eq:R-transform}
\tilde{R}(t) = Q(t)R(t)Q^\intercal(0).
\end{equation}
Our goal now can be stated as finding two different control Hamiltonians in the same frame that satisfy three conditions.
\begin{enumerate}
    \item The same final gate is produced for both control Hamiltonians, $\tilde{U}_c(T) = U_c(T)$.
    \item A geometric phase is traded for a dynamical one via Eqs.~\eqref{eq:geom-phase-change}-\eqref{eq:dyn-phase-change}.
    \item The relevant filter function(s) produced is (are) the same for both control Hamiltonians.
\end{enumerate}
We can satisfy the first condition by requiring that $V(0) = V(T) = \mathbbm{1}$ and likewise for $Q$. The second condition can be satisfied by finding a $Q$ such that a geometric evolution $R(t)$ is related to a dynamical evolution $\tilde{R}(t)$ by Eq.~\eqref{eq:R-transform}. The third condition can be satisfied if, for a given noise source $\delta_q$, the integrands in Eq.~\eqref{eq:R-freq} are equal; i.e., combining with Eq.~\eqref{eq:R-transform}, it suffices that
\begin{equation}\label{eq:equal-FF}
Q^\intercal(t) \bm{\chi}_q\left[\tilde{\bm{h}}_c(t)\right]
= \bm{\chi}_q\left[\bm{h}_c(t)\right].
\end{equation}
Note that the sensitivity vector $\bm{\chi}_q$ has a fixed functional dependence on its control input. (If we were making a frame transformation, this term would be transformed in the same way as the control Hamiltonian.) This reflects the fact that the underlying noise mechanism is fixed by the physics of the device, and is not under the control of the user. As an example, if we have some control field $h_1(t)$ with an error model $h_1(t) \rightarrow h_1(t)(1+\delta_1(t))$, then any other choice of that control field must have the same noise dependence: $\tilde{h}_1(t) \rightarrow \tilde{h}_1(t)(1+\delta_1(t))$. More generally, the sensitivity vector is simply evaluated as a function of the new control input
\begin{equation}
    \bm{\tilde{h}_c}(t) = Q(t)\bm{h_c}(t)+\bm{h_{Q}}(t), \,\,
    h_{Q,i}(t) \equiv \tr(-i Q \dot{Q}^T \Lambda_i)/N,
\end{equation}
where $\bm{\Lambda}$ is the $(N^2-1)$-dimensional vector of generators of $\mathfrak{su}(N)$ isomorphic to $\bm{\sigma}$. In conjunction with Eq.~\eqref{eq:equal-FF}, this yields the condition
\begin{gather}
    Q^\intercal(t) \bm{a}_q + Q^\intercal(t) M_q(t) (Q(t) \bm{h_c}(t)+\bm{h_{Q}}(t)) \nonumber\\ \label{eq:equal-FF2}
    = \bm{a_q} + M_q(t) \bm{h_c}(t).
\end{gather}
If we can find a $Q(t)$ that satisfies Eq.~\eqref{eq:equal-FF2}, we will have two Hamiltonians that result in identical gates and filter functions but with different phase types. This simple fact is the crux of this paper. We present in Fig.~\ref{fig:overview} a schematic diagram that summarizes our main result.

While the existence of a solution to the nonlinear Eq.~\eqref{eq:equal-FF2} is not obvious, we simplify by taking the more restrictive condition that the first (second) term on the lhs must separately equal the first (second) term on the rhs. Thus, one should choose $Q(t)$ such that i) $\bm{a_q}$ is an eigenvector of $Q(t)$, ii) $[Q(t),M_q(t)]=0$, and iii) $\bm{h_Q}(t)$ is in the null space of $M_q(t)$. Parameterizing as $Q(t)=\mathcal{T}\exp{\int \bm{\omega}(t)\cdot\bm{\Lambda} \mathrm{d}t}$, these conditions become i) $\omega_i(t) = 0$ if $\bm{a_q} \in \text{Col}\, \Lambda_i$, ii) $\left[\bm{\omega}(t)\cdot \bm{\Lambda},M_q(t)\right] = 0$, and iii) $\bm{\omega}(t) \in \text{null} \, M_q(t)$. In practice, it is typically easy to satisfy these conditions.

\section{EXAMPLES}
\label{sec:examples}
\subsection{Abelian case}
To illustrate, consider a single qubit under a generic $\mathfrak{su}(2)$ control Hamiltonian
\begin{equation}\label{eq:h1q}
\bm{h_c}(t) = \frac{1}{2}\begin{pmatrix}
\Omega(t)\cos(\varphi(t))\\
\Omega(t)\sin(\varphi(t))\\
\Delta(t)
\end{pmatrix}.
\end{equation}
This form is pertinent to a variety of qubit implementations such as superconducting qubits \cite{Koch_2007}, quantum dot spin qubits \cite{Laucht_2017}, and NMR qubits \cite{Gershenfeld_1997} to name a few, corresponding to the rotating-wave approximation for a two-level system driven by an oscillating field with amplitude $\Omega$ at a carrier frequency detuned from resonance by $\Delta$, and with phase $\varphi$. Suppose that this qubit is subject to independent additive fluctuations in the resonance frequency, $\Delta \rightarrow \Delta + \delta_\Delta$, and in the phase, $\varphi \rightarrow \varphi + \delta_\varphi$, as well as multiplicative amplitude noise, $\Omega \rightarrow \Omega (1+\delta_\Omega)$, i.e., in terms of Eq.~\eqref{eq:chi},
\begin{align}
\bm{a_\Delta} &= 
\frac{1}{2}\bm{\hat{z}}, \quad 
M_\Delta = 0,
\\
\bm{a_{\varphi}} &= \bm{0}, \quad
M_\varphi = \frac{1}{2}\left(\bm{\hat{y}}\bm{\hat{x}}^T-\bm{\hat{x}}\bm{\hat{y}}^T\right),
\\
\bm{a_{\Omega}} &= \bm{0}, \quad
M_\Omega = \frac{1}{2}\left(\bm{\hat{x}}\bm{\hat{x}}^T + \bm{\hat{y}}\bm{\hat{y}}^T\right).
\end{align}
Note that this noise model encompasses most, if not all, scenarios treated in the literature of nonadiabatic geometric gates.

The three flavors of Eq.~\eqref{eq:equal-FF2} corresponding to $q=\Delta, \varphi, \Omega$ are all satisfied by choosing $Q(t) = \mathrm{e}^{\nu(t)\Lambda_z}$ (and hence $\bm{h_Q} = \dot{\nu}(t)\bm{\hat{z}}/2$) where $\Lambda_z$ is the $z$-rotation generator in $\mathfrak{so}(3)$ and $\nu(t)$ is any function satisfying $\nu(0) = \nu(T) = 0$. Thus, for any geometric gate produced by a particular choice of $\Omega(t)$, $\varphi(t)$, and $\Delta(t)$ in Eq.~\eqref{eq:h1q}, one can implement the same gate with identical robustness by using the modified control
\begin{equation}
\label{eq:Ham-family}
\bm{\tilde{h}_c}(t) = \frac{1}{2}\begin{pmatrix}
\Omega(t)\cos(\varphi(t)+\nu(t))\\
\Omega(t)\sin(\varphi(t)+\nu(t))\\
\Delta(t)+\dot{\nu}(t)
\end{pmatrix}
\end{equation}
and the free parameter $\nu(t)$ allows a way to tune the nature of the Lewis-Riesenfeld phase as indicated in Eqs.~\eqref{eq:geom-phase-change} and \eqref{eq:dyn-phase-change}.

\begin{figure}
    \raggedright
    \includegraphics[scale=0.40]{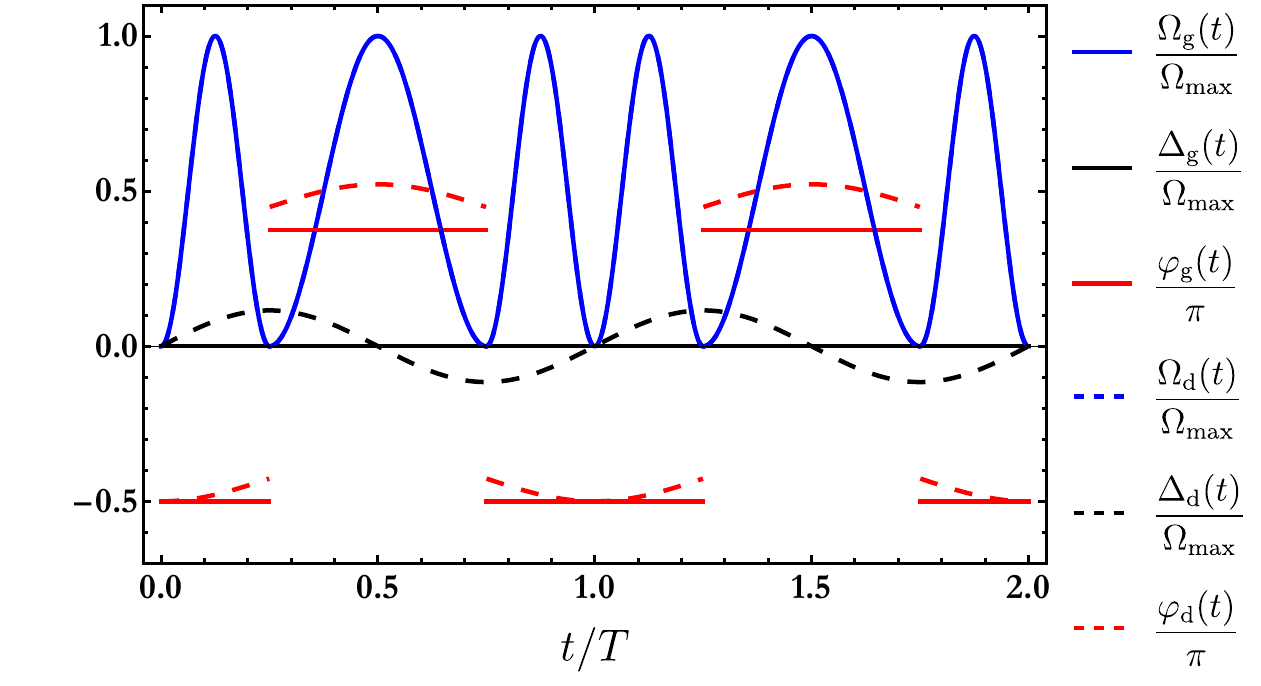}
    \caption{A plot of the control parameters that generate an $X_{\frac{\pi}{2}}$ gate. The subscript ``g" (``d") denote the control parameters that generate a geometric (dynamical) gate. The values are normalized by $\Omega_{max}$ which denote the maximum value of $\Omega_{g}(t)$. Note that $\Omega_{g}(t) = \Omega_{d}(t)$ and that $\Delta_{d}(t)$ is non-trivial.}
    \label{fig:equiv-ctrl}
\end{figure}

\begin{figure}
    \raggedright
    \includegraphics[scale=0.46]{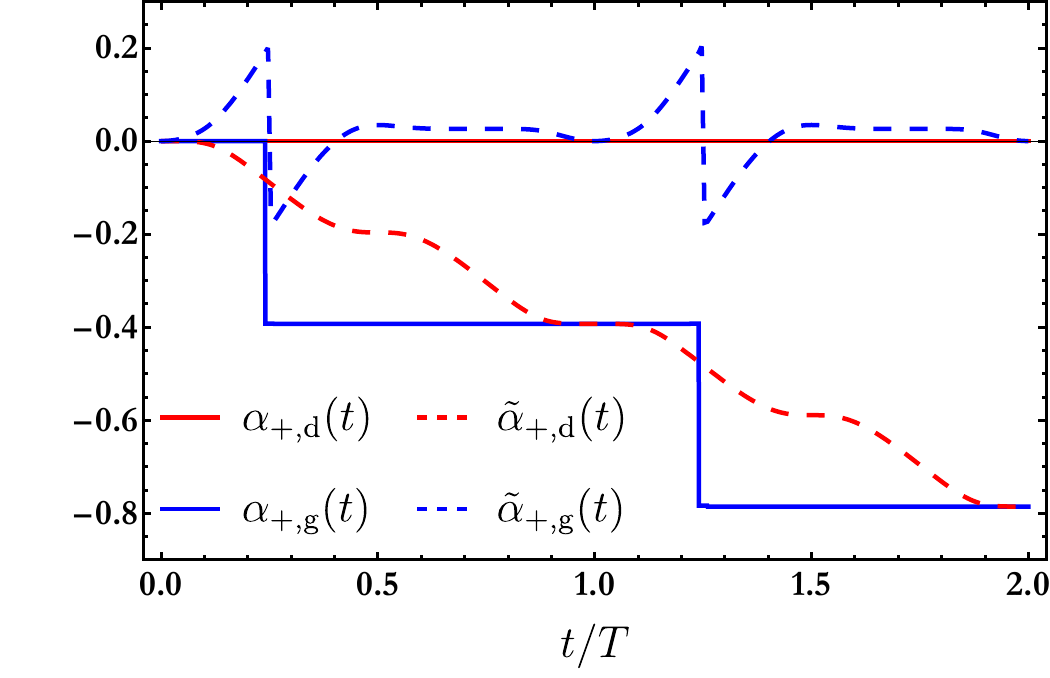}
    \caption{A comparison of the geometric and dynamical phases generated by the state $\ket{\phi_+(t)}$. The variables with (without) tilde correspond to the dynamical (geometric) $X_{\frac{\pi}{2}}$ gate.}
    \label{fig:phase}
\end{figure}

\begin{figure}
    \raggedright
    \includegraphics[scale=0.48]{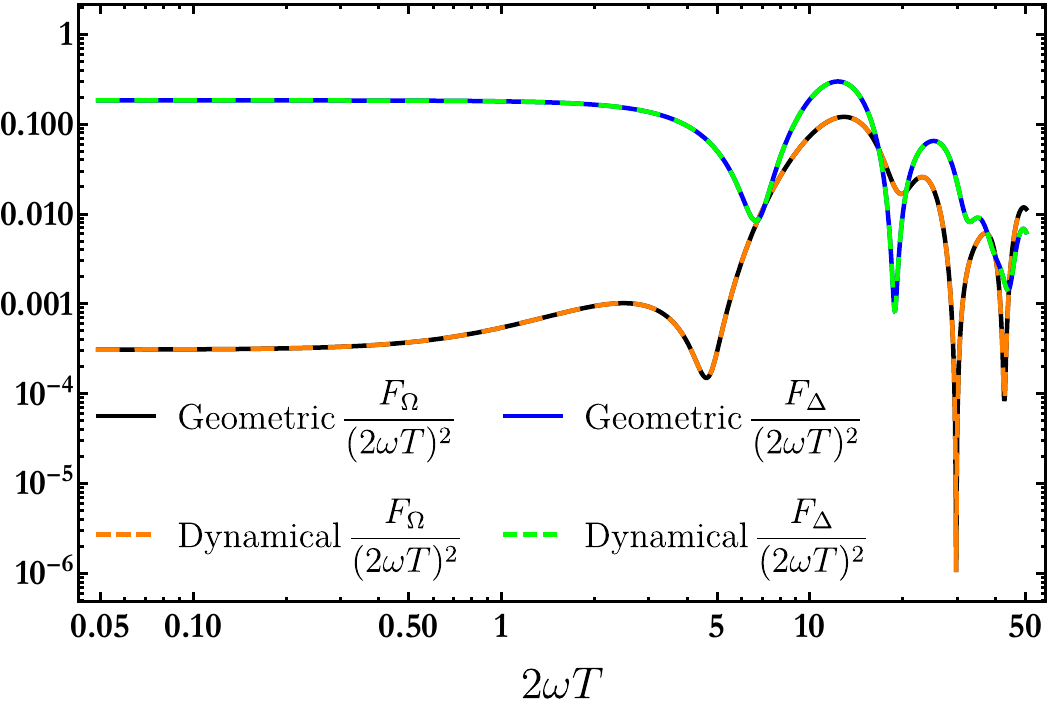}
    \caption{A comparison of the geometric and dynamical $X_{\frac{\pi}{2}}$ gate filter functions for additive dephasing and multiplicative amplitude noise when $\Omega_{max} = 1$. We verify that the two control Hamiltonians produce the same filter functions.}
    \label{fig:ff-comparison}
\end{figure}

\subsubsection{Orange-slice scheme}
We now provide explicit examples of matching the noise sensitivity of a geometric gate with an equivalent dynamical gate. We begin by considering the Abelian geometric gate proposed in Ref.~\cite{Zhao_2017}. The dynamical invariant eigenstate traces out an orange-slice path along the Bloch sphere (see Fig.~\ref{fig:path}), and the geometric phase is equivalent to half the enclosed area \cite{Sjoqvist_2016}. We present in Eqs.~\eqref{eq:s1}-\eqref{eq:s3} the corresponding control constraints in terms of the Hamiltonian of Eq.~\eqref{eq:h1q} with $\Delta(t) = 0$:
\begin{align}
\label{eq:s1}
&t\in\left[0,T_1\right]&     \int_{0}^{T_1} \Omega\mathrm{d}t &= \theta&        &\varphi = \eta - \frac{\pi}{2}, \\
\label{eq:s2}
&t\in\left[T_1,T_2\right]&   \int_{T_1}^{T_2}\Omega\mathrm{d}t &= \pi&          &\varphi = \eta +\gamma +\frac{\pi}{2}, \\
\label{eq:s3}
&t\in\left[T_2,T\right]&     \int_{T_2}^{T}\Omega\mathrm{d}t &= \pi-\theta&     &\varphi = \eta - \frac{\pi}{2}.
\end{align}
We denote the generated evolution operator by $U_0(t) = e^{i\gamma \bm{n}\cdot\bm{\sigma}}$, where $\bm{n} = \left(\sin\theta\cos\varphi,\sin\theta\sin\varphi,\cos\theta\right)$. We can produce a two-part composite gate that suppresses additive dephasing $(\Delta \rightarrow \Delta + \delta_\Delta)$ and multiplicative amplitude noise $(\Omega \rightarrow \Omega + \delta_\Omega \Omega)$ by applying the same evolution twice \cite{Chen_2018}: $U(2T) = U_0^2(T)$. We target a geometric $X_{\frac{\pi}{2}}$ gate which can be achieved by setting $\gamma = -\frac{\pi}{8}, \theta = \frac{\pi}{2},$ and $\eta=0$. We assume the pulse shape $\Omega(t) = \sin^2\left(\pi t/\tau\right)$ where $\tau$ is the length of the relevant time interval. To generate its purely dynamical equivalent, we use the modified Hamiltonian of Eq.~\eqref{eq:Ham-family} with an arbitrary choice of $\nu(t) = c \sin^2\left(\pi t/T\right)$, numerically tuning $c$ until the geometric phase is zero, which occurs at $c \approx 0.461875$.  We present in Fig.~\ref{fig:equiv-ctrl} a plot of the control parameters for both geometric and dynamical $X_{\frac{\pi}{2}}$ gates. We use Eqs.~\eqref{eq:geometric-phase} and \eqref{eq:dynamical-phase} to verify that the modified Hamiltonian produces a purely dynamical gate. The dynamical invariant eigenvectors $\ket{\phi_\pm(t)}$ are determined using the inverse engineering scheme in Ref.~\cite{Chen_2011}:
\begin{gather}
    \ket{\phi_+(t)} = \begin{pmatrix}
    \cos\left(\frac{\gamma(t)}{2}\right) \exp\left(-i\beta(t)\right)\\
    \sin\left(\frac{\gamma(t)}{2}\right)
    \end{pmatrix}, \\
    \ket{\phi_-(t)} = \begin{pmatrix}
    \sin\left(\frac{\gamma(t)}{2}\right) \\
    -\cos\left(\frac{\gamma(t)}{2}\right)\exp\left(i\beta(t)\right)
    \end{pmatrix}.
\end{gather}
Here the parameters $\gamma$ and $\beta$ obey the coupled differential equations:
\begin{gather}
    \dot{\gamma} = -\Omega \sin\left(\beta-\varphi\right),\\
    \dot{\beta} = \Delta - \Omega \cot\gamma \cos\left(\beta-\varphi\right).
\end{gather}
We set the boundary conditions so that $\gamma(0) = \frac{\pi}{2}$ and $\beta(0) = 0$, which corresponds to an eigenvector of $X_{\frac{\pi}{2}}$.
The effect of this evolution on $\ket{\phi_+(t)}$ is shown in Fig.~\ref{fig:path}. Finally, a comparison of the geometric and dynamical phases for both gates is shown in Fig.~\ref{fig:phase} and their corresponding filter functions for dephasing and amplitude noise in Fig.~\ref{fig:ff-comparison}.

\begin{figure*}
    \centering
    \includegraphics[scale=0.7]{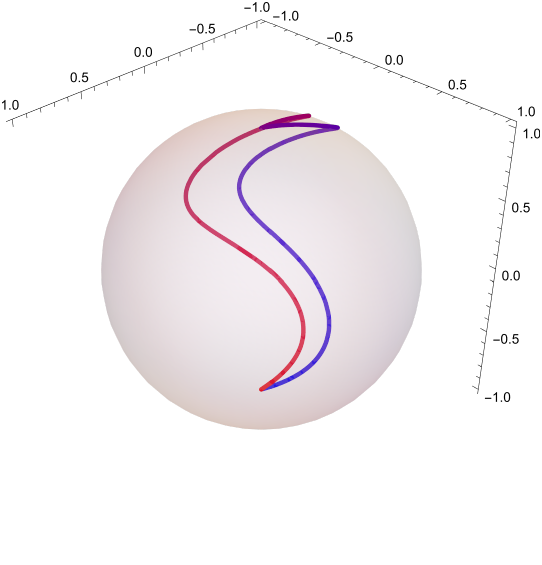}
    \includegraphics[scale=0.7]{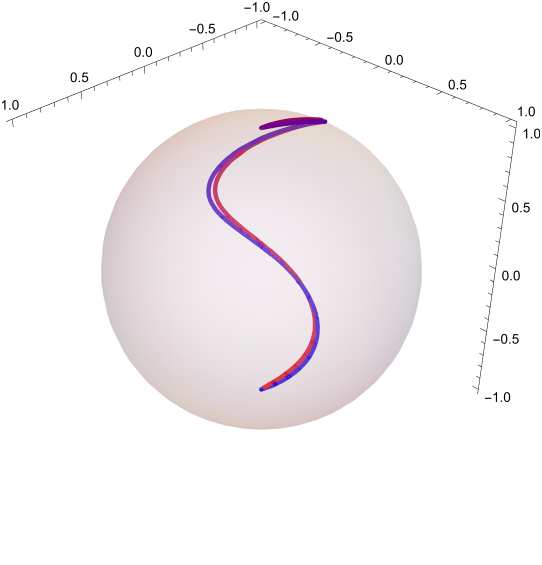}
    \caption{An illustration of a dynamical invariant eigenvector's evolution along the Bloch sphere for the geometric T-gate (LEFT) and the dynamical T-gate (RIGHT). The path orientation is determined by the color gradient which begins with red and ends with blue. Just like in the previous example, we find that $\ket{\phi_+(t)}$ traces out a loop with nonzero area in the geometric case and zero area in the dynamical case.}
    \label{fig:path-oct}
\end{figure*}

\subsubsection{Inverse engineering with optimal control}
\begin{figure}
    \raggedright
    \includegraphics[scale=0.4]{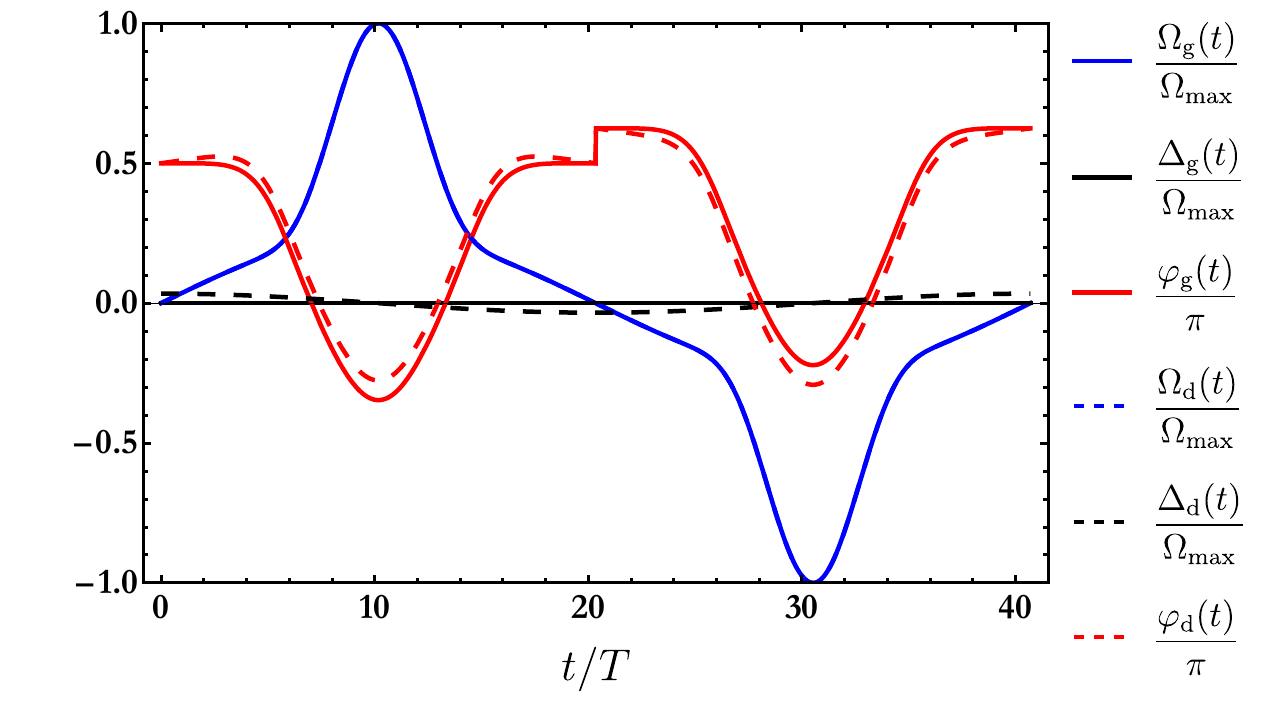}
    \caption{A plot of the control parameters that generate a $Z_{\frac{\pi}{4}}$ or T-gate. The subscript ``g" (``d") denote the control parameters that generate a geometric (dynamical) gate. The values are normalized by $\Omega_{max}$ which denote the maximum value of $\Omega(t)$. We note again that $\Omega_{g}(t) = \Omega_{d}(t)$ and that $\Delta_{d}(t)$ is non-trivial.}
    \label{fig:equiv-ctrl-oct}
\end{figure}

\begin{figure}
    \raggedright
    \includegraphics[scale=0.46]{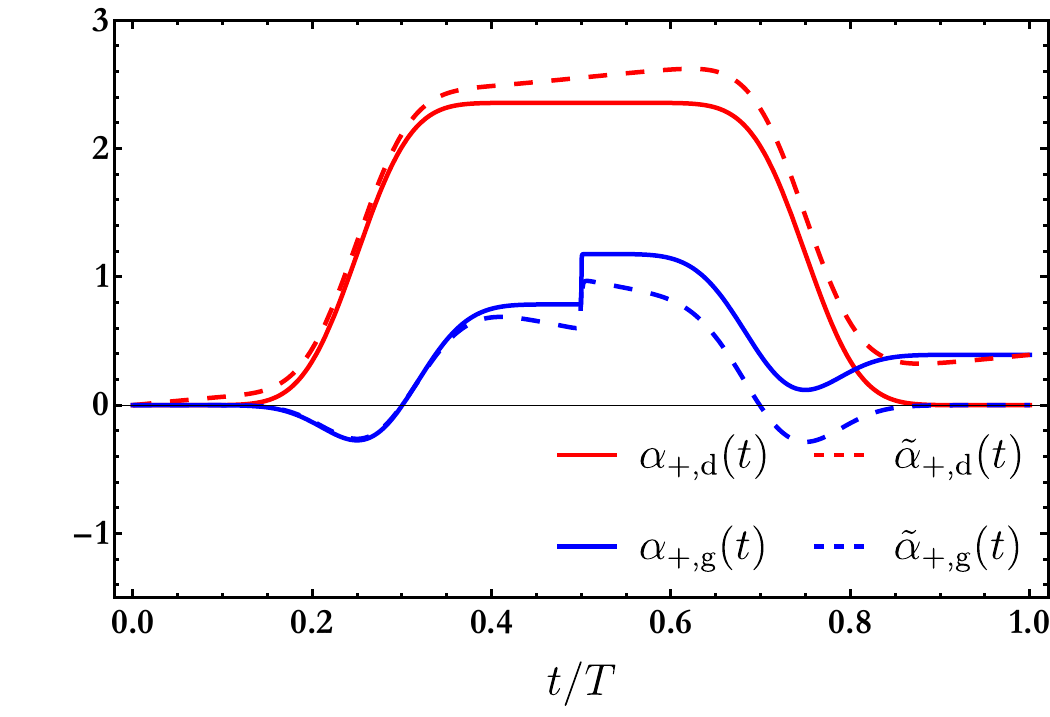}
    \caption{A comparison of the geometric and dynamical phases generated by the state $\ket{\phi_+(t)}$. The variables with (without) tilde correspond to the dynamical (geometric) T-gate.}
    \label{fig:phase-oct}
\end{figure}

\begin{figure}
    \raggedright
    \includegraphics[scale=0.48]{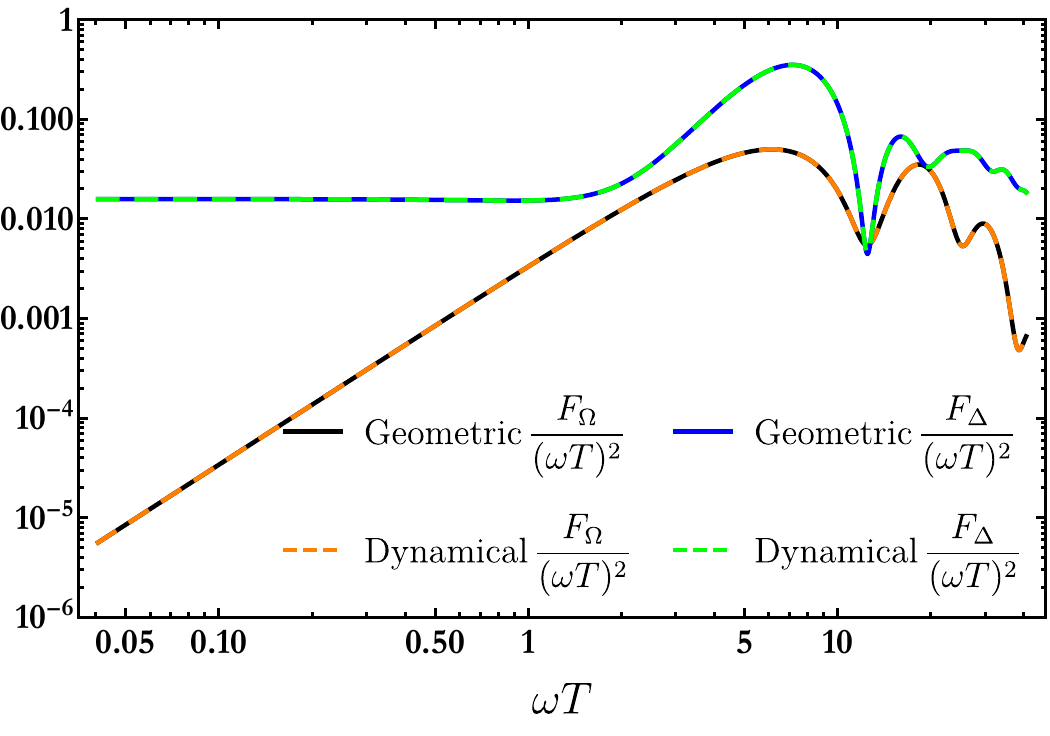}
    \caption{A comparison of the geometric and dynamical T-gate filter functions for additive dephasing and multiplicative amplitude noise when $\Omega_{max} = 1$. We verify that the two control Hamiltonians produce the same filter functions.}
    \label{fig:ff-comparison-oct}
\end{figure}

Next, we consider the case of a nonadiabatic Abelian geometric gate that is produced using Hamiltonian inverse engineering and optimal control theory \cite{J-Xu_2020}. Suppose that we target a T-gate ($Z_{\frac{\pi}{4}}$) as in Ref. \cite{J-Xu_2020}. The optimized inverse-engineered control parameters are given by
\begin{gather}
    \Omega = -\frac{\dot{\gamma}}{\sin\left(\beta - \varphi \right)}, \\
    \varphi = \beta - \arctan \left( \frac{\dot{\gamma}}{\dot{\beta}}\cot \gamma \right), \\
    \Delta = 0,
\end{gather}
which depend on the piecewise-defined functions $\gamma(t)$ and $\beta(t)$ that satisfy
\begin{align}
    t \in \left[0,T/2\right]: &\quad \gamma(t) = \pi \sin^2(\pi t/T),\\
                              &\quad \beta(t) = -\frac{4}{3}\cos\left(\frac{\pi}{2}\cos\left(\frac{2\pi t}{T}\right)\right),\\
    t \in \left[T/2,T\right]: &\quad \gamma(t) = \pi \sin^2(\pi t/T),\\
                              &\quad \beta(t) = -\frac{4}{3}\cos\left(\frac{\pi}{2}\cos\left(\frac{2\pi t}{T}\right)\right) + \frac{\pi}{8}.
\end{align}
The gate time is $T = \frac{\sqrt{17}\pi^2}{\Omega_{\text{max}}}$, where $\Omega_{\text{max}}$ is the maximum value of $\Omega(t)$. To generate its purely dynamical equivalent, we again use the modified Hamiltonian of Eq.~\eqref{eq:Ham-family} with an arbitrary choice of $\nu(t) = c \sin\left(\frac{2\pi t}{T}\right)$, numerically tuning $c$ until the geometric phase is zero, which occurs at $c \approx 0.220530$. We set the boundary conditions so that $\gamma(0) = \beta(0) = \pi$ which corresponds to an eigenvector of $Z_{\frac{\pi}{4}}$. The effect of this evolution on $\ket{\phi_+(t)}$ is shown in Fig.~\ref{fig:path-oct}. We present in Fig.~\ref{fig:equiv-ctrl-oct} a plot of the geometric and the dynamical T-gate's control parameters. Finally, a comparison of the geometric and dynamical phases for both gates is shown in Fig.~\ref{fig:phase-oct} and their corresponding filter functions for dephasing and amplitude noise in Fig.~\ref{fig:ff-comparison-oct}.

\subsection{Non-Abelian case}
We now extend this treatment to the non-Abelian case. Unlike the Abelian case, where ensuring that the dynamical phase is zero at the final time is a constraint on the geometric pulse design, non-Abelian geometric quantum computing typically encodes the computational basis in an energetically degenerate subspace of the full Hilbert space such that any dynamical phase is either automatically zero at all times or can be treated as a global phase factor. We generalize our previous framework and denote the eigenvectors of $I(t)$ by $\ket{\phi_{n;a}(t)}$ where $a \in \{ 1, 2, \ldots d_n \}$  labels the orthonormal basis vectors of a $d_n$-fold degenerate subspace corresponding to the $n^\text{th}$ eigenvalue. The propagator of Eq.~\eqref{eq:U_c} generalizes to \cite{Mostafazadeh_2001}
\begin{equation}
\label{eq:U_c-deg}
U(t) = \sum_{n} \sum_{a,b=1}^{d_n} u_{n;ab}(t) \ket{\phi_{n;a}(t)}\bra{\phi_{n;b}(0)},
\end{equation}
where the eigenstates accumulate a non-Abelian phase $u_{n}(t)$ given by
\begin{gather}
\label{eq:lewis-riesenfeld-phase-deg}
u_n(t) = \mathcal{T} e^{i\int_{0}^{t} \mathcal{A}_n(t^\prime) + \mathcal{E}_n(t^\prime) \mathrm{d}t^\prime}, \\
\label{eq:geometric-phase-deg}
\mathcal{A}_{n;ab}(t) = \mel**{\phi_{n;a}(t)}{i \partial_t}{\phi_{n;b}(t)}, \\
\label{eq:dynamical-phase-deg}
\mathcal{E}_{n;ab}(t) = -\mel**{\phi_{n;a}(t)}{H(t)}{\phi_{n;b}(t)}.
\end{gather}
Thus, if we again consider the effect of changing the Hamiltonian as expressed by a quantum canonical transformation (again, without loss of generality) via time-dependent unitary $V$, we get expressions for the changes in the geometric and dynamical components of the Lewis-Riesenfeld phase similar to Eqs.~\eqref{eq:geom-phase-change} and \eqref{eq:dyn-phase-change}:
\begin{gather}
\label{eq:geom-phase-change-nonAbel}
\tilde{\mathcal{A}}_{n;ab}(t) = \mathcal{A}_{n;ab}(t) + \mel**{\phi_{n;a}(t)}{i V^\dagger \dot{V}}{\phi_{n;b}(t)}, \\
\label{eq:dyn-phase-change-nonAbel}
\tilde{\mathcal{E}}_{n;ab}(t) = \mathcal{E}_{n;ab}(t) - \mel**{\phi_{n;a}(t)}{i V^\dagger \dot{V}}{\phi_{n;b}(t)}.
\end{gather}

We note that the dynamical and geometric contributions to the phase are easily separable in the Abelian case, as in Eqs.~\eqref{eq:geom-phase-change} and \eqref{eq:dyn-phase-change}. In the non-Abelian case, the gate accumulates matrix-valued dynamical and geometric phase components at each time step, as seen in Eq.~\eqref{eq:lewis-riesenfeld-phase-deg}, which generally do not commute. Thus, the inseparability of the phase's time-ordered integral can lead to nontrivial dynamical contributions even if the integral of $\mathcal{E}(t)$ in Eq.~\eqref{eq:dynamical-phase-deg} is zero. This is why purely geometric non-Abelian gates are typically defined to have $\mathcal{E}(t) = 0$ within the computational basis \cite{Sjoqvist_2012,Sjoqvist_2016}. Therefore, to illustrate that a non-Abelian gate is no longer purely geometric, it is sufficient to show that the transformation $V$ produces a nontrivial $\tilde{\mathcal{E}}$ in Eq.~\eqref{eq:dyn-phase-change-nonAbel}. Specifically, it is sufficient to show that $\tilde{\mathcal{E}}_{n;ab}(t) \neq 0$ where $a$ and $b$ index the computational basis states. 

On the other hand, it is possible to simultaneously diagonalize both matrices in special cases where $\left[\mathcal{A}(t),\mathcal{E}(t)\right] = 0$ which allows the decoupling of the geometric and dynamical phase contributions (for example, in adiabatic non-Abelian geometric gates \cite{Zanardi_1999}). In such cases, it is again straightforward to separate and tune the two types of phase.

\subsubsection{Nonadiabatic case}
As an example, consider a three-level system in a $\Lambda$ configuration where the states $\ket{0}$ and $\ket{1}$ are coupled to an excited state $\ket{e}$ \cite{Sjoqvist_2012}. The $k \leftrightarrow e$ transition $(k = 0,1)$ is separately driven by a laser pulse with fixed polarization and frequency. Following our notation in Eq.~\eqref{eq:H}, the system-laser interaction is described by the following rotating-frame Hamiltonian belonging to an $\mathfrak{su}(3)$ algebra:
\begin{equation}
\label{eq:holonomic-H}
\bm{h_c}(t)= \begin{pmatrix}
0\\
0\\
\frac{\Delta_0-\Delta_1}{2}\\
\Omega\left(t\right)\cos\left(\frac{\varphi}{2}\right) \sin\left(\frac{\theta}{2}\right)\\
\Omega\left(t\right)\sin\left(\frac{\varphi}{2}\right) \sin\left(\frac{\theta}{2}\right)\\
-\Omega\left(t\right)\cos\left(\frac{\varphi}{2}\right)\cos\left(\frac{\theta}{2}\right)\\
\Omega\left(t\right)\sin\left(\frac{\varphi}{2}\right)\cos\left(\frac{\theta}{2}\right)\\
\frac{\Delta_0+\Delta_1}{2\sqrt{3}}
\end{pmatrix}.
\end{equation}
Here $\theta$ and $\varphi$ are fixed angles that describe the relative strength and relative phase of the $k \leftrightarrow e$ transitions, $\Delta_k$ are detunings that can be independently varied, $\Omega(t)$ describes the pulse amplitude envelope, and $\bm{\sigma}$ is chosen to comprise the Gell-Mann matrices. If we impose the constraint that $\int_{0}^{T}\Omega(t)\mathrm{d}t = \pi$ and drive the qubit at resonance $(\Delta_k=0)$, the evolution produces a purely geometric gate that, when projected in the computational space spanned by $\{\ket{0}, \ket{1}\}$, yields \cite{Sjoqvist_2012}
\begin{equation}
\label{eq:holonomic-gate}
\text{proj}_{\{\ket{0}, \ket{1}\}}\left[U(T)\right] = \begin{pmatrix}
\cos \theta & e^{-i\varphi}\sin \theta\\
e^{i\varphi}\sin \theta & -\cos \theta
\end{pmatrix}.
\end{equation}
It is possible to generate any single-qubit operation by applying Eq.~\eqref{eq:holonomic-gate} with different values of $\theta$ and $\varphi$.

Suppose that this qubit is subject to independent additive fluctuations in the laser detunings, $\Delta_k \rightarrow \Delta_k + \delta_{\Delta_k}$, in their relative strength, $\theta \rightarrow \theta + \delta_\theta$, and in their relative phase, $\varphi \rightarrow \varphi + \delta_\varphi$, as well as multiplicative amplitude noise, $\Omega \rightarrow \Omega (1+\delta_\Omega)$. Then, in terms of Eq.~\eqref{eq:chi}, we have
\begin{align}
&\left(\bm{a_{\Delta_{k}}}\right)_i = (-1)^k \pi \delta_{i,3} + \frac{\pi}{\sqrt{3}} \delta_{i,8} \, , \quad
M_{\Delta_k} = 0,\\
&\bm{a_{\Omega}} = \bm{0}, \quad
M_{\Omega} = E_{4,4}+E_{5,5}+E_{6,6}+E_{7,7},\\
&\bm{a_{\theta}} = \bm{0}, \quad
M_{\theta} = \frac{1}{2} \left(E_{5,7}-E_{7,5}+E_{6,4}-E_{4,6}\right),\\
&\bm{a_{\varphi}} = \bm{0}, \quad
M_{\varphi} = \frac{1}{2} \left(E_{5,4}-E_{4,5}+E_{6,7}-E_{7,6}\right),
\end{align}
where $E_{i,j}$ is a square matrix with value $1$ at position $(i,j)$ and zeros elsewhere \cite{Pfeifer_2003}. It is straightforward to verify that the transformation $Q(t) = \mathrm{e}^{\nu(t) \Lambda_3}$, where the $\Lambda_i$ are the adjoint representations of the Gell-Mann matrices \footnote{We distinguish the adjoint representation of a group which is defined in Eq.~\eqref{eq:adjoint-rep} from the adjoint representation of a Lie algebra which can be calculated using the structure constants of the algebra $f_{ijk}$ obeying $[\sigma_i,\sigma_j]=\sum_k i f_{ijk} \sigma_k$ as $[\text{ad}(\sigma_i)]_{jk} = -i f_{ijk}$.}, uniquely satisfies all the previously specified criteria. (If any one of these noise sources is irrelevant, there is more freedom in the transformation.) Thus, for any non-Abelian gate produced by a particular choice of $\theta, \varphi$, and $\Omega(t)$ in Eq.~\eqref{eq:holonomic-gate}, one can implement the same gate with identical robustness using the modified control 
\begin{equation}
\label{eq:nonAbel-family}
\bm{\tilde{h}_c}(t)= \begin{pmatrix}
0\\
0\\
\frac{\Delta_0-\Delta_1+\nu^\prime(t)}{2}\\
\Omega\left(t\right)\cos\left(\frac{\varphi+\nu\left(t\right)}{2}\right) \sin\left(\frac{\theta}{2}\right)\\
\Omega\left(t\right)\sin\left(\frac{\varphi+\nu\left(t\right)}{2}\right) \sin\left(\frac{\theta}{2}\right)\\
-\Omega\left(t\right)\cos\left(\frac{\varphi+\nu\left(t\right)}{2}\right)\cos\left(\frac{\theta}{2}\right)\\
\Omega\left(t\right)\sin\left(\frac{\varphi+\nu\left(t\right)}{2}\right)\cos\left(\frac{\theta}{2}\right)\\
\frac{\Delta_0+\Delta_1}{2\sqrt{3}}
\end{pmatrix}
\end{equation}
where the free parameter $\nu(t)$ breaks the degeneracy of an equal-detuning setting, and similar to the Abelian case, provides a way to tune the nature of the Lewis-Riesenfeld phase as indicated in Eqs.~\eqref{eq:geom-phase-change-nonAbel} and \eqref{eq:dyn-phase-change-nonAbel}.

As previously mentioned, we need only show that our transformation yields $\tilde{\mathcal{E}}_{n;ab}(t) \neq 0$ within the computational subspace to guarantee that the gate is no longer purely geometric. Since $H(t)$ commutes with itself at all times, we can calculate the resulting evolution operator $U(t)$ analytically
\begin{equation}
    U(t) = \\ \exp\left[-i \overline{\Omega}(t)
   \begin{pmatrix}
0\\
0\\
0\\
\cos\left(\frac{\varphi}{2}\right) \sin\left(\frac{\theta}{2}\right)\\
\sin\left(\frac{\varphi}{2}\right) \sin\left(\frac{\theta}{2}\right)\\
-\cos\left(\frac{\varphi}{2}\right)\cos\left(\frac{\theta}{2}\right)\\
\sin\left(\frac{\varphi}{2}\right)\cos\left(\frac{\theta}{2}\right)\\
0
\end{pmatrix}\cdot\bm{\sigma}\right],
\end{equation}
where we define $\overline{\Omega}(t) \equiv \int_{0}^{t} \Omega(s)\mathrm{d}s$. We note that a dynamical invariant can be constructed by using the cyclic states of $U(T)$ as its eigenbasis \cite{Mostafazadeh_2001}. To proceed, we first compute the eigenvectors and eigenvalues of $U(t)$:
\begin{align}
    \lambda_1(t) =& 1  &\ket{\lambda_1(t)}& = \begin{pmatrix}e^{-i\frac{\varphi}{2}}\cos\left(\frac{\theta}{2}\right)\\e^{i\frac{\varphi}{2}}\sin\left(\frac{\theta}{2}\right)\\0\end{pmatrix}, \\
    \lambda_2(t) =& e^{-i\overline{\Omega}(t)}  &\ket{\lambda_2(t)}& = \frac{1}{\sqrt{2}} \begin{pmatrix}e^{-i\frac{\varphi}{2}}\sin\left(\frac{\theta}{2}\right)\\-e^{i\frac{\varphi}{2}}\cos\left(\frac{\theta}{2}\right)\\1\end{pmatrix}, \\
    \lambda_3(t) =& e^{i\overline{\Omega}(t)}  &\ket{\lambda_3(t)}& = \frac{1}{\sqrt{2}} \begin{pmatrix}-e^{-i\frac{\varphi}{2}}\sin\left(\frac{\theta}{2}\right)\\e^{i\frac{\varphi}{2}}\cos\left(\frac{\theta}{2}\right)\\1\end{pmatrix}. 
\end{align}
The cyclic states $\ket{\phi_i(t)}$ of $U(T)$ are linear combinations of $\ket{\lambda_i}$ up to global time-dependent phase that we choose so that $\ket{\phi_i(0)}=\ket{\phi_i(T)}$:
\begin{gather}
    \ket{\phi_1(t)} = \ket{\lambda_1(t)},\\ \qquad \ket{\phi_2(t)} = e^{i\overline{\Omega}(t)}\frac{\lambda_2(t)\ket{\lambda_2(t)}-\lambda_3(t)\ket{\lambda_3(t)}}{\sqrt{2}},\\
    \qquad \ket{\phi_3(t)} = e^{i\overline{\Omega}(t)}\frac{\lambda_2(t)\ket{\lambda_2(t)}+\lambda_3(t)\ket{\lambda_3(t)}}{\sqrt{2}}.
\end{gather}
Consequently, we can define a dynamical invariant as $I(t) = \sum_i c_i \ket{\phi_i(t)}\bra{\phi_i(t)}$, where the $c_i$ are arbitrary constants. Using the eigenvectors of $I(t)$, we can calculate the change in the dynamical phase contribution under the transformation $V = e^{-i \frac{\nu(t)}{2}\lambda_3}$ (or, equivalently, by $Q = e^{\nu(t)\Lambda_3}$ in the adjoint representation) using Eq.~\eqref{eq:dyn-phase-change-nonAbel}:
\begin{widetext}
\begin{equation}
    \tilde{\mathcal{E}}(t) = 
        \frac{1}{4}\begin{pmatrix}
        -2\cos(\theta)\nu'(t) & -\sin\left(\theta\right)\nu'(t)\left(1+e^{i2\overline{\Omega}(t)}\right) & -\sin\left(\theta\right)\nu'(t)\left(1-e^{i2\overline{\Omega}(t)}\right) \\
        -\sin\left(\theta\right)\nu'(t)\left(1+e^{-i2\overline{\Omega}(t)}\right) & 2\cos(\theta)\cos^2(\overline{\Omega}(t))\nu'(t) & -4\Omega(t) - i\cos\left(\theta\right)\sin\left(2\overline{\Omega}(t)\right)\nu'(t)\\
        -\sin\left(\theta\right)\nu'(t)\left(1-e^{-i2\overline{\Omega}(t)}\right) & -4\Omega(t) + i\cos\left(\theta\right)\sin\left(2\overline{\Omega}(t)\right)\nu'(t) & 2\cos(\theta)\sin^2\left(\overline{\Omega}(t)\right)\nu'(t)
        \end{pmatrix}
\end{equation}
\end{widetext}
Since $\tilde{\mathcal{E}}(t)$ is nontrivial in the computational subspace, then the gate $\tilde{U}(T)$ must not be purely geometric by definition, though its filter function is the same as $U(T)$.

\subsubsection{Adiabatic case}
\label{subsubsec:adiabatic}
We next consider the case of an adiabatic non-Abelian geometric gate. Specifically, we consider a four-level system with three ground or metastable states coupled to a single excited state as in Refs.~\cite{Duan_2001,Zhang_2015}. The system is controlled using three distinctly polarized and resonantly driven lasers that, in the rotating frame, yield the control Hamiltonian
\begin{equation}
    \bm{\tilde{h}_c}(t)= \begin{pmatrix} 0\\ 0\\ 0\\ \Omega(t)\cos\varphi(t) \sin\theta(t)\\ \Omega(t)\sin\varphi(t)\sin\theta(t)\\ \Omega(t)\cos\theta(t) \end{pmatrix},
\end{equation}
where $\theta$ controls the relative strength between the lasers, $\varphi$ controls their relative phases, and $\bm{\sigma}$ is chosen to comprise of the $\mathfrak{so}(4)$ generators
\begin{align}
    e_1& = i(E_{2,1}-E_{1,2}), & e_2& = i(E_{3,1}-E_{1,3}), &\nonumber\\
    e_3& = i(E_{3,2}-E_{2,3}), & e_4& = E_{4,1} + E_{1,4}, &\\
    e_5& = E_{4,2} + E_{2,4}, & e_6& = E_{4,3} + E_{3,4}.\nonumber
\end{align}
We assume for simplicity that all laser magnitudes are constant through the evolution and are sufficiently large to ensure adiabaticity. The control parameters $\theta$ and $\varphi$ are then tuned cyclically so that at the gate time $t = T$ we have $H(0) = H(T)$. The dynamics of the system can be described using the eigenvectors of $H(t)$:
\begin{align}
\lambda =& 0 &\ket{\lambda_{1,0}}& = \begin{pmatrix}\cos\theta(t) \cos\varphi(t)\\ \cos\theta(t) \sin\varphi(t)\\ -\sin\theta(t)\\ 0 \end{pmatrix},\\
\lambda =& 0  &\ket{\lambda_{1,1}}& = \begin{pmatrix} -\sin\varphi(t)\\ \cos\varphi(t)\\ 0\\ 0 \end{pmatrix},\\
\lambda =& -\Omega(t)  &\ket{\lambda_2}& = \frac{1}{\sqrt{2}}\begin{pmatrix} \cos\varphi(t)\sin\theta(t)\\ \sin\varphi(t)\sin\theta(t)\\ \cos\theta(t)\\ -1 \end{pmatrix},\\
\lambda =& \Omega(t)  &\ket{\lambda_3}& = \frac{1}{\sqrt{2}}\begin{pmatrix} \cos\varphi(t)\sin\theta(t)\\ \sin\varphi(t)\sin\theta(t)\\ \cos\theta(t)\\ 1 \end{pmatrix},
\end{align}
with $\ket{\lambda_{1,0}}$ and $\ket{\lambda_{1,1}}$ spanning the energetically degenerate computational basis. In this basis, $\mathcal{E}_{n;ab}(t) = \varepsilon_n(t)\delta_{ab}$, which consequently decouples $\mathcal{E}$ and $\mathcal{A}$ in Eq.~\eqref{eq:lewis-riesenfeld-phase-deg}. Thus, the adiabatic evolution operator is given by
\begin{equation}
    U(t) = \sum_{n=1}^{3} \sum_{a,b=0}^{1} u_{n;ab} \ket{\lambda_{n;a}(t)}\bra{\lambda_{n;b}(t)},
\end{equation}
where
\begin{gather}
    u_n(t) = e^{-i\int_0^t \varepsilon_n(t)} \mathcal{T} e^{\int_0^t \mathcal{A}_n(t')\mathrm{d}t'},\\
\mathcal{A}_{n;ab}(t) = \mel**{\lambda_{n;a}(t)}{i\partial_t}{\lambda_{n;b}(t)}. 
\end{gather}
As a result, any accumulated dynamical phase in the computational basis can be treated as a global phase factor. Therefore, the nontrivial effects of the evolution are completely due to the non-Abelian geometric phase. 

Suppose that this system is subject to independent additive fluctuations in the lasers' relative strength, $\theta \rightarrow \theta + \delta_\theta$, in their relative phase, $\varphi \rightarrow \varphi + \delta_\varphi$, as well as multiplicative amplitude noise, $\Omega \rightarrow \Omega(1+\delta_\Omega)$. Then, in terms of Eq.~\eqref{eq:chi}, we have
\begin{align}
&\bm{a_{\Omega}} = \bm{0}, \quad
&M_{\Omega} =& E_{4,4}+E_{5,5}+E_{6,6},\\
&\bm{a_{\theta}} = \bm{0}, \quad
&M_{\theta} =& -\tan\theta(t)E_{6,6},\\
&\bm{a_{\varphi}} = \bm{0}, \quad
&M_{\varphi} =& \frac{1}{2} \left(E_{5,4}-E_{4,5}\right).
\end{align}
It can be easily verified that the transformation $V = e^{-i \nu(t) e_1}$ (or, equivalently, by $Q = e^{2\nu(t)\mathrm{ad}(e_1)}$) satisfies the conditions we outlined in the main text, where $\mathrm{ad}(e_1)$ denotes the adjoint representation of $e_1$. We can calculate the corresponding change in the dynamical phase contribution using Eq.~\eqref{eq:dyn-phase-change-nonAbel}:
\begin{widetext}
\begin{equation}
    \tilde{\mathcal{E}}(t) = 
        \begin{pmatrix}
        0 & i\cos\left(\theta(t)\right)\nu'(t) & 0 & 0\\
        -i\cos\left(\theta(t)\right)\nu'(t) & 0 & -\frac{i\sin\left(\theta(t)\right)\nu'(t)}{\sqrt{2}} & -\frac{i\sin\left(\theta(t)\right)\nu'(t)}{\sqrt{2}}\\
        0 & \frac{i\sin\left(\theta(t)\right)\nu'(t)}{\sqrt{2}} & -\Omega(t) & 0\\
        0 & \frac{i\sin\left(\theta(t)\right)\nu'(t)}{\sqrt{2}} & 0 & \Omega(t)
        \end{pmatrix},
\end{equation}
\end{widetext}
with the computational subspace located in the upper $2\times 2$ block. We again see that $\tilde{\mathcal{E}}(t)$ is nontrivial in the computational subspace which indicates that the gate $\tilde{U}(T)$ is not purely geometric even though its filter function is the same as $U(T)$. We further note that this transformation yields a control Hamiltonian with nondegenerate energy levels. Thus, the geometric and dynamical components of the phase integral in Eq.~\eqref{eq:lewis-riesenfeld-phase-deg} are no longer decoupled, which is in contrast with the non-Abelian version of the gate.

\section{DISCUSSION}
\label{sec:discussion}

\begin{figure}
    \centering
    \includegraphics[scale=0.42]{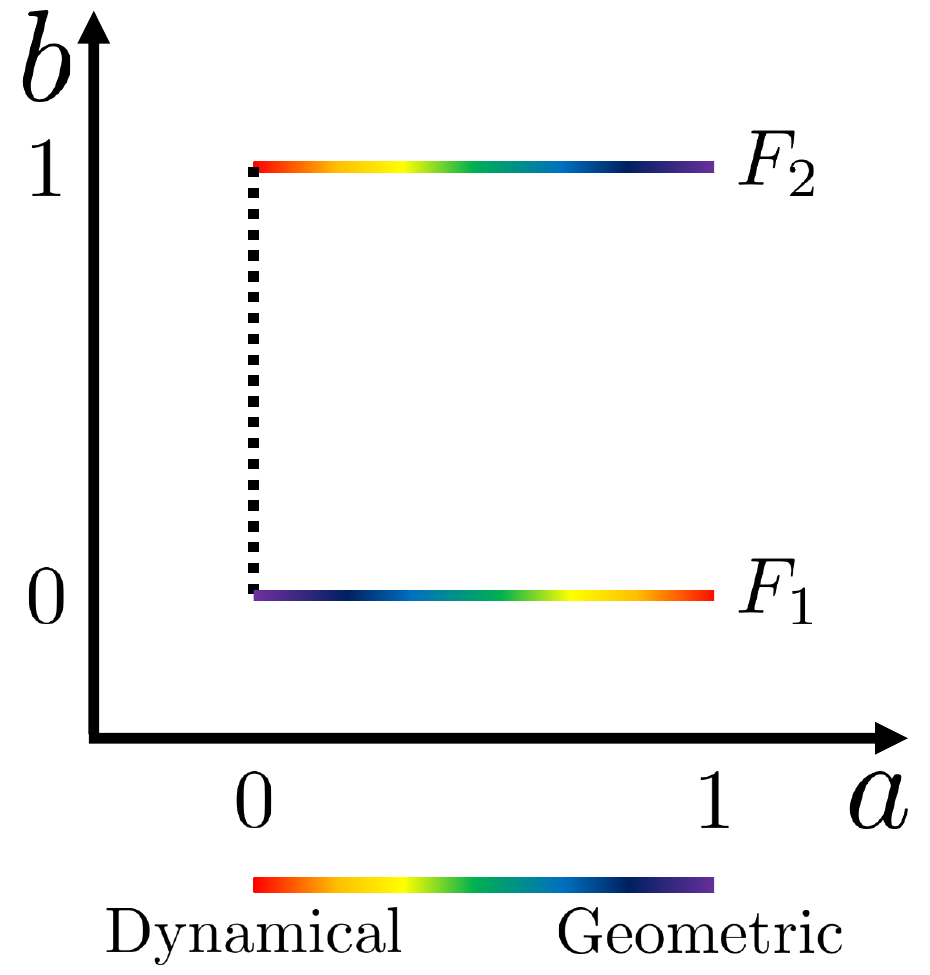}
    \caption{A schematic illustration demonstrating how control constraints can give rise to preferential phase robustness. Here we consider the set of control Hamiltonians described by Eq.~\eqref{eq:parameterized-H}. The colored horizontal lines represent families of control Hamiltonians that preserve the filter function, $F_1$ and $F_2$ respectively. The color gradient indicates the gate's phase type, which can range from purely dynamical to purely geometric. The dotted line represents a subset of the control space that is physically accessible in a given experiment as a result of strict constraints (generally this dotted line will broaden into an extended region of the plane). In this case, the constraint prohibits one from traversing the horizontal lines, so one would obtain a different noise sensitivity (i.e., filter function) for a geometric gate versus a dynamical gate.}
    \label{fig:discussion}
\end{figure}

We demonstrated in Sec.~\ref{sec:examples} that in many experimentally relevant scenarios there exist families of solutions to Eq.~\eqref{eq:equal-FF2} which implies that the notion of noise robustness is independent of phase type. This contradicts the expectation that geometric gates are superior to dynamical gates since it is always possible in principle to find two Hamiltonians that produce the same gate and noise sensitivity but with polar opposite phase types. In other words, there is nothing particularly special about geometric gates when it comes to robustness. At first glance, this may seem to contradict the significant evidence in the literature that supports the robustness claim for geometric gates. It is crucial to note, however, that the solution set of Eq.~\eqref{eq:equal-FF2} is only non-trivial when the control Hamiltonian is not severely constrained. Here constraints refer to the physical limitations of a specific qubit implementation such as control parameter bounds, only two-axis control, or bandwidth limitations. Depending on the error model in consideration and the severity of the constraints, there can be scenarios where the only solution to Eq.~\eqref{eq:equal-FF2} is a trivial one ($Q(t) = \mathbbm{1}$). This clearly happens, for example, with an unusual error model of $\bm{a_q}=0$ and $M_q=\mathbbm{1}$ in Eq.~\eqref{eq:chi}. Likewise, geometric gates naturally emerge as superior in the particular case of a strictly two-axis control Hamiltonian with static multiplicative amplitude error \cite{Ichikawa_2012}. In such special cases, the correspondence between filter function and phase type is unique, i.e., phase preference emerges. 

We can illustrate how phase preference emerges using Fig.~\ref{fig:discussion}. Generally one has many tunable parameters in the control Hamiltonian (e.g., even for only a single time-dependent control field, one has the value of the field over each infinitesimal time step). However, for the sake of being able to sketch an illustration, consider a control Hamiltonian with only two free parameters,
\begin{gather}
    \bm{h_c}(t;a,b) = Q(t)\left[b \bm{h_g}(t) + (1-b) \bm{h_d}(t)\right] + \bm{h_Q}(t), \\
    Q(t) = \mathcal{T}\exp{a\int \left[b\bm{\omega_g}(t) + (1-b)\bm{\omega_d}(t)\right]\cdot\bm{\Lambda}\mathrm{d}t}\label{eq:parameterized-H},
\end{gather}
where $\bm{h_g}$ ($\bm{h_d}$) denotes a specific physically accessible control Hamiltonian that produces a particular target gate geometrically (dynamically), $\bm{\omega_g}$ ($\bm{\omega_d}$) denotes a rotation axis vector that determines an operator $Q$ which solves Eq.~\eqref{eq:equal-FF2} such that it produces the same target gate with an identical filter function but a different phase type (assuming such a solution exists for the relevant error model, as in the examples of Sec.~\ref{sec:examples}), and $a, b$ parameterize a continuous deformation between these four specific points in the control space.

The horizontal lines of Fig.~\ref{fig:discussion} are sets of control fields that all yield the same filter function (labeled $F_1$ and $F_2$ in the figure). The color gradient indicates the phase type across these lines which can range from purely dynamical to purely geometric. However, physical constraints may only allow access to some subarea of the $a-b$ plane. For example, in a severely constrained case, one may only have access to Hamiltonians with $a=0$ in Eq.~\eqref{eq:parameterized-H}, indicated by the dotted line in Fig.~\ref{fig:discussion}. So, although in this paper we have shown that a geometric gate generally has a dynamical equivalent with equal noise sensitivity (and vice versa), a strict control constraint could prohibit one from accessing these equivalent controls in practice. As depicted in Fig.~\ref{fig:discussion}, one would then observe different noise sensitivities for different phase types. 

A notable example where this behavior is observed is in Ref.~\cite{Zhu_2005}. Here the authors considered the control Hamiltonian in Eq.~\eqref{eq:h1q} with the constraint that $\varphi = \varphi_0 t$ and $\Omega, \Delta,$ and $\varphi_0$ are constants. In addition, they assumed that $\Omega$ and $\Delta$ are subject to multiplicative noise. By further imposing the restriction that $\varphi = \left[\Delta \pm \sqrt{\Delta^2 - \eta(\Omega^2 + \Delta^2)}\right]/\eta$ and $\Delta = \Omega \sqrt{\eta/(1-\eta)} + \Delta_0$ where $\eta$ and $\Delta_0$ are constants, it is possible for them to switch between a geometric and a dynamical gate. However, it can be easily verified that the constraints they take do not permit the control to be changed as prescribed in Eq.~\eqref{eq:Ham-family}. Thus, when they change the gate's phase type (move up/down the dotted line), they also changed the gate's filter function. In their case, they found that geometric gates performed better than dynamical gates. In the situation of Ref.~\cite{Blais_2003}, where a strictly two-axis, piecewise constant control scheme with additive noise on both axes was considered, the constraints again preclude moving along the horizontal lines of Fig.~\ref{fig:discussion} but in this case it is dynamical gates that were found to perform better than geometric gates.

Thus, our result can be used to reconcile seemingly contradictory results in the literature regarding the robustness of geometric vs.~dynamical gates, in that the studies that reached different conclusions also imposed different control constraints. We can interpret the apparent superiority of either phase type as a consequence of the constraints ruling out a solution to Eq.~\eqref{eq:equal-FF2} since we have shown in Sec.~\ref{sec:examples} that they are generically equivalent in the absence of constraints. Determining in which scenario dynamical gates or geometric gates are superior can only be done on a case-by-case basis as it is determined by the noise model as well as the particular constraints. For example, Ref.~\cite{Colmenar_2022} examined the Hamiltonian in Eq.~\eqref{eq:h1q} under the constraint that $\Delta$ is constant. It was found that any gate that is robust against static multiplicative amplitude noise $(\delta_\Omega)$ as well as static additive or multiplicative detuning noise $(\delta_\Delta)$ is necessarily geometric. A more general investigation of constraint types that favor geometric gates remains an interesting and challenging open question. We emphasize, however, that constraints do not necessarily favor geometric gates. If we only consider static detuning noise, it becomes possible to find robust dynamical gates as well. 

Realizing equivalent Hamiltonians may require degrees of freedom in the control to be present in one that are not present in the other. For example, dynamic control of the detuning, $\Delta(t)$, is necessary to produce the dynamical gates seen in Fig.~\ref{fig:equiv-ctrl} and \ref{fig:equiv-ctrl-oct}. While controlling the detuning is not entirely common, this level of control has already been achieved in superconducting qubits \cite{Lucero_2010} and in quantum dot charge qubits \cite{Kim_2015}. 

Finally, we clarify that although we have only shown the existence of nontrivial solutions to Eq.~\eqref{eq:equal-FF2} for a restricted set of coherent error models, our analysis encompasses all coherent error models that we are aware of having been considered in the literature. In addition, our findings can also be extended to account for dissipative processes. We have verified through Lindblad master equation simulations that the equivalent geometric and dynamical gates are identically affected by dephasing and relaxation. This behavior is expected since, by construction, the two Hamiltonians produce gates with the same duration and filter function.

\section{Conclusion}
In summary, we examine the broadband noise-resilience of geometric and dynamical gates using filter functions and show that in the absence of control constraints there is generally no intrinsic advantage for one or the other -- for any control Hamiltonian producing a geometric gate one can find a different control Hamiltonian that produces a completely equivalent dynamical gate in the same frame. We illustrate this explicitly in a one-qubit scenario for both the Abelian and non-Abelian case. Our analysis applies to both adiabatic and non-adiabatic gates and does not impose any speed restriction on the control. We discuss how the presence of control constraints can give rise to preferential phase robustness and reconcile apparently contradictory claims in the current literature regarding the robustness of geometric gates. Since geometric gates are not inherently more robust than dynamical gates, then the utility of geometric quantum computing becomes a matter of the specific experimental control constraints, and a broad categorization of which types of constraints favor geometric gates remains an important open question.

RKLC and JPK acknowledge support from the National Science Foundation under Grant No. 1915064, and UG from the Army Research Office under Grant No.~W911NF-17-1-0287.

\bibliography{geometric-and-dynamical-gate-robustness}
\bibliographystyle{apsrev4-2}

\end{document}